\documentclass[english,twocolumn,aps,prl, superscriptaddress,showpacs, amsmath]{revtex4-1}

\usepackage{graphicx}
\usepackage{dcolumn}
\usepackage[toc,page]{appendix}
\usepackage{bm}
\usepackage{xcolor}
\usepackage{float}
\usepackage[bottom]{footmisc}
\usepackage{lipsum}
\usepackage{background}
\backgroundsetup{
  position=current page.north,
  angle=0,
  nodeanchor=north,
  vshift=-2mm,
  opacity=1,
  scale=2,
  contents= Draft
}

\begin{document}


\title{A magnetic-field-free superconducting mixed state}
\title{Mixed superconducting state without applied magnetic field}

\author{Alex Khanukov}
\affiliation{ Department of Physics, Bar Ilan University, Ramat Gan 5290002, Israel}
\affiliation{Institute of Nanotechnology and Advanced Materials, Bar-Ilan University, Ramat Gan 5290002, Israel.}
\author{Itay Mangel}
\affiliation{ Department of Physics, Technion Israel Institute of Technology, Haifa 3200003, Israel}
\author{Shai Wissberg}
\affiliation{ Department of Physics, Bar Ilan University, Ramat Gan 5290002, Israel}
\affiliation{Institute of Nanotechnology and Advanced Materials, Bar-Ilan University, Ramat Gan 5290002, Israel.}
\author{Amit Keren}
\email{phkeren@technion.ac.il}
\affiliation{ Department of Physics, Technion Israel Institute of Technology, Haifa 3200003, Israel}
\author{Beena Kalisky}
\email{beena@biu.ac.il}
\affiliation{ Department of Physics, Bar Ilan University, Ramat Gan 5290002, Israel}
\affiliation{Institute of Nanotechnology and Advanced Materials, Bar-Ilan University, Ramat Gan 5290002, Israel.}


\begin{abstract}

 A superconducting (SC) mixed state occurs in type-II superconductors where the upper critical field $H_{c2}$ is higher than the thermodynamic critical field $H_c$. When an applied field is in between these fields, the free energy depends weakly on the order parameter which therefore can be small (SC state) or zero (normal state) at different parts of the sample. In this paper we demonstrate how a normal state along a line traversing a superconductor can be turned on and off externally in zero field. The concept is based on a long, current-carrying excitation coil, piercing a ring-shaped superconductor. The ring experiences zero field, but the vector potential produced by the coil generates a circular current that destroys superconductivity along a radial line starting at preexisting nucleation points in the sample. Unlike the destruction of superconductivity with magnetic field, the vector potential method is reversible and reproducible; full superconductivity is recovered upon removing the current from the coil, and different cooldowns yield the same normal lines. We suggest potential applications of this magnetic-field-free mixed state.

\end{abstract}

\maketitle


\section{\label{sec:level1}Introduction}

When a long current-carrying excitation coil (EC) pierces a superconducting (SC) ring, as in Fig.~\ref{fig:setup}(a), current is expected to flow in the ring and produce a magnetic field that expels the flux from the ring's hole. The ring's current density $\bf{j}$, is determined by the vector potential $\bf{A}$ produced by the EC, the current in the ring, and by the complex order parameter $\Psi = \psi({\bf r}) {e^{i\phi ({\bf{r}})}}$, with $\psi({\bf r})\ge 0$, according to
\begin{equation}
{\bf{j}} = \rho_s \left( {\frac{\Phi_0}{2\pi} \nabla \phi  - {\bf{A}}} \right)
\label{JtoAandPhi}
\end{equation}
where  $\Phi_0$ is the superconducting flux quanta,
\begin{equation}
{\rho _s} = \frac{{ \psi  ^2{{e^*}^2}}}{{m^*}},
\end{equation} 
and $e^*$ and $m^*$ are the carriers charge and mass respectively. $\psi ^2$ is often interpreted as a measure of the superconducting carrier density with a maximum value $ \psi_0 ^2$ \cite{GAVISH2021132767}. If at some point on the SC ring the current density exceeds the critical value $j_c$ of the ring's material, the order parameter drops to zero at this point \cite{GAVISH2021132767}. 
In this setup, the SC ring does not experience any applied magnetic field and the current in the ring is set by the Faraday effect when the current in the EC is turned on. 

For a system with rotation symmetry, it is expected that the current will be strongest at the inner rim of the ring \cite{GAVISH2021132767}. If this current exceeds $j_c$, the order parameter will diminish first at the inner rim, followed by the full destruction of superconductivity as the EC current increases. This should lead to a radial propagation of a normal state front as illustrated in Fig.~\ref{fig:setup}(b).

In addition, the ring does experience fields generated by its own supercurrents. These self-fields, in principle, could generate vortices where the order parameter is destroyed locally. If and when vortices are present, they are described by $\nabla \phi = 1/|\bf{r}-\bf{r}_v| $ where $\bf{r}_v$ is the location of the vortex. For a system with strict rotation symmetry vortices are not allowed. But, in the real world, ring-shaped materials with perfect rotation symmetry are hard to come by. In parallel, theory for situations lacking rotation symmetry is difficult. Therefore, the initial motivation for this paper was to examine the behavior of a realistic ring experimentally with a local magnetic mapping technique: a scanning superconducting quantum interference device (SQUID). 

Surprisingly, we found that in the non-rotational-symmetric case, the superconducting order parameter is destroyed along radial lines emanating from defects leading to a new type of mixed state. In fact, no propagating front was detected what so ever. This stands in strong contrast to the expectation in the rotationally symmetric case. The destruction of the order parameter is reversible; when the coil current is turned off, the radial destruction of superconductivity disappears. The behavior of the system is also reproducible between different cool downs. This is a very different situation from dendrite growth which occurs when superconductivity is destroyed by an external field.

\section{\label{sec:level1} Experiment}
\begin{figure}[H]
\includegraphics[width=0.5\textwidth]{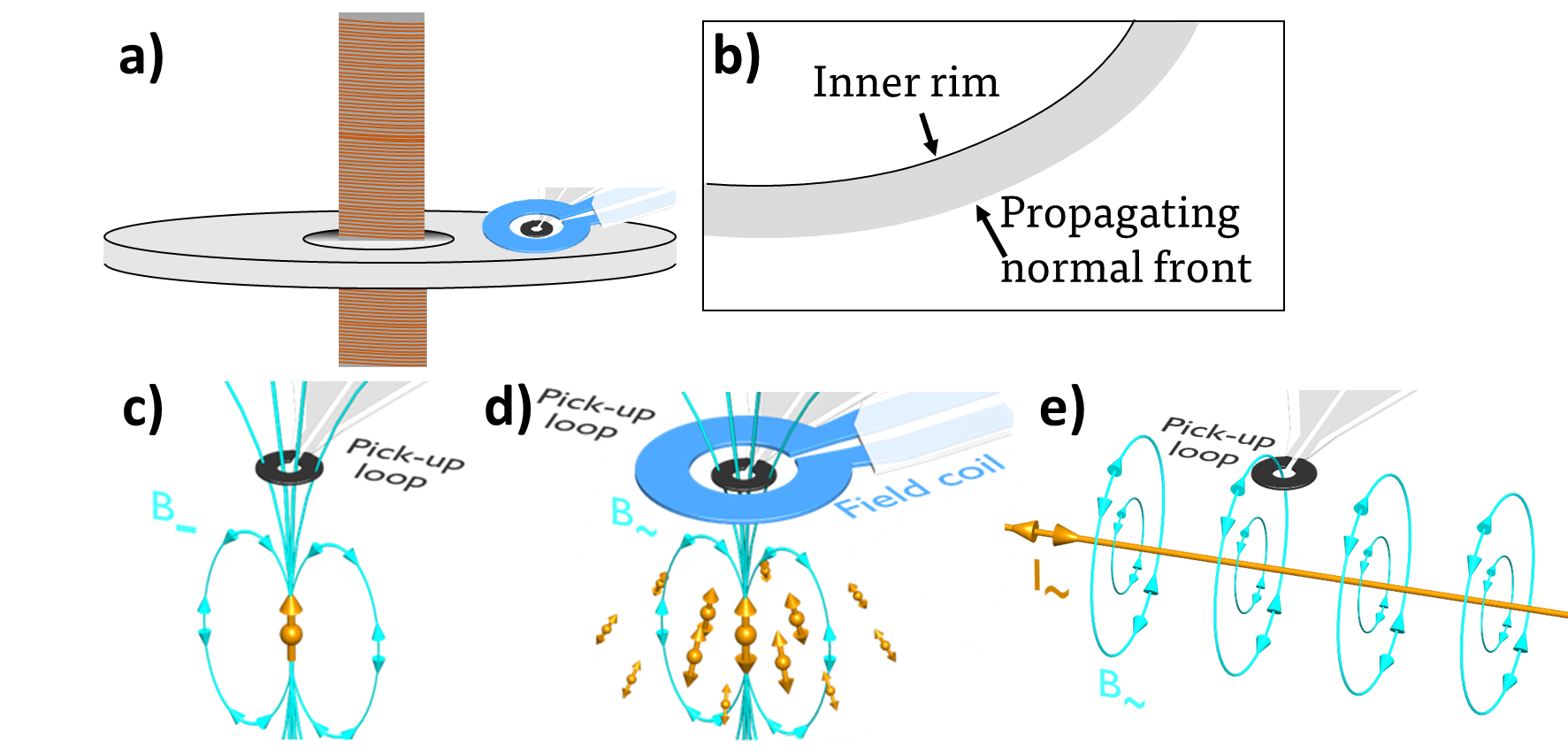}
\caption{\label{fig:setup} \textbf{Scanning SQUID imaging of a ring threaded by an "infinite" excitation coil }. (a) Illustration of the measurement configuration. The ring is fixed and is threaded by a long coil and is scanned by the SQUID. (b) Illustration of a normal front propagating from the inner rim of the ring. (c)-(e) Illustration of scanning SQUID detection modes. (c) Static magnetic flux. (d) Diamagnetic response, the flux generated by supercurrents flowing in the sample as a response to a locally applied ac magnetic field, by a field coil on the SQUID chip. (e) Flux generated by a line of current. This mode is used to record the flow of supercurrents in the sample as a response to an ac current in the EC. }
\end{figure}

The key feature required for our experiment is a long excitation coil compared to the superconducting ring dimensions, passing at its center, as shown in Fig.~\ref{fig:setup}(a). The coil is considered as infinitely long with zero magnetic field outside but with a finite vector potential. The quality of this approximation is discussed in the Appendix. In practice, the coil is $60$~mm long, with a diameter of $0.7$~mm, $6$ layers, a total of $7200$ windings, and is made of NbTi. The superconducting rings were made of $8$~nm MoSi film \cite{Keren_2022}, with an outer diameter of $5$~mm and inner diameter of $1$~mm. The films were grown on a ring-shaped silicone wafer with a $300$-nm silicone oxide layer. We used a scanning SQUID susceptometer to simultaneously record the local susceptibility, current flow, and local static magnetism \cite{doi:10.1063/1.2932341, doi:10.1146/annurev-conmatphys-031620-104226}. 

\begin{figure}[H]
\includegraphics[width=0.5\textwidth]{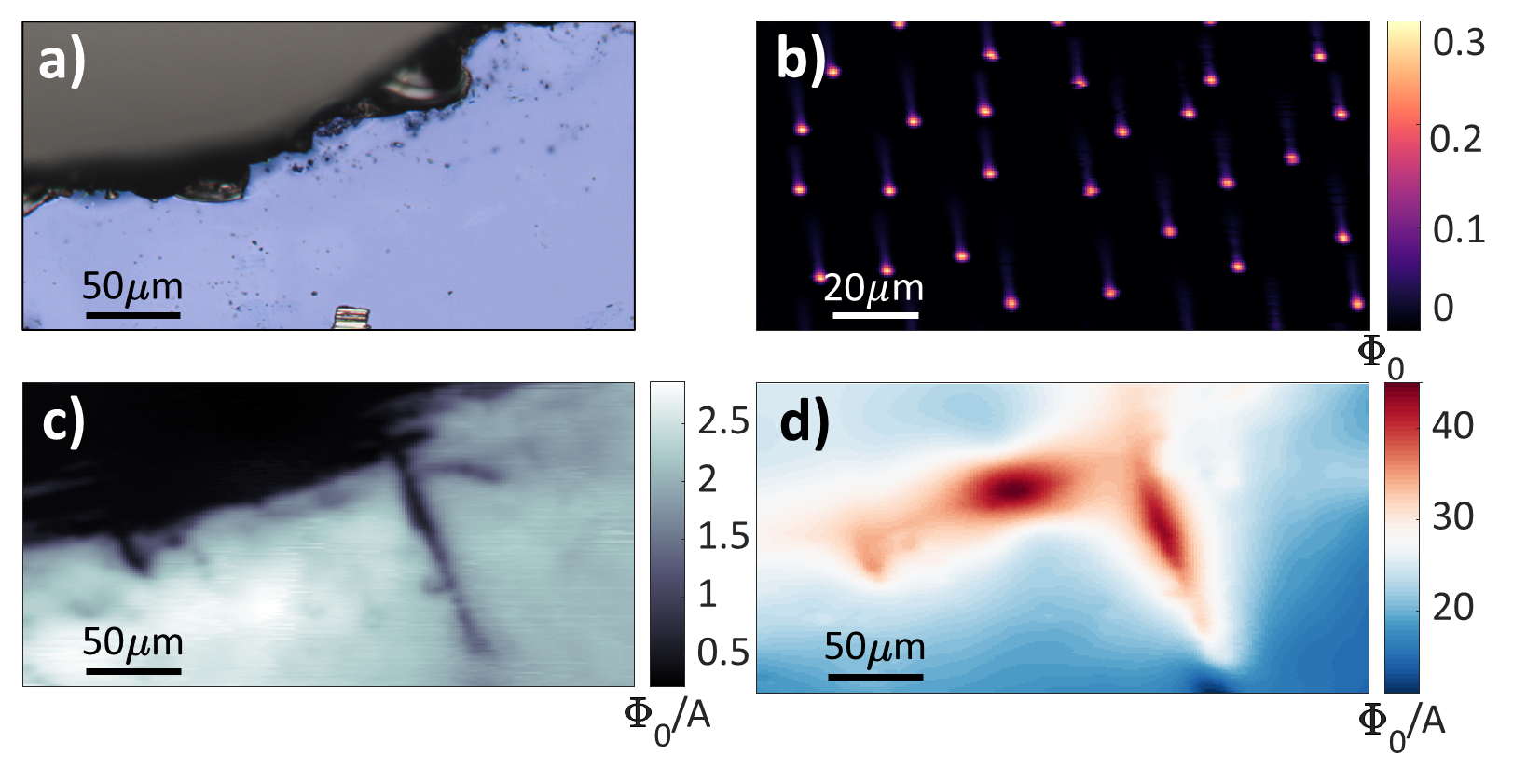}
\caption{\label{fig:scanned} \textbf{Scanning SQUID data}. (a) Optical image of the inner rim of the ring. (b) Image of the static magnetic flux, showing
superconducting vortices on the ring at a field of $81.9$~mG. (c) Susceptibility image of the area shown in (a), taken in the presence of $I=0.093$~mA in the EC (d) Flux generated by supercurrents in the ring, flowing as a response to $I=0.093$~mA in the EC.}
\end{figure}

The experimental configuration is illustrated in Fig.~\ref{fig:setup}(a). The ring-coil pairs are fixed in place, while the SQUID sensor is scanning near the ring's surface. An optical picture of the inner rim of the ring is presented in Fig.~\ref{fig:scanned}(a). Three measurement modes, illustrated in Figs.~\ref{fig:setup}(c) -~\ref{fig:setup}(e), were operated simultaneously: \par
(A) The static magnetic landscape is mapped by rastering the SQUID over the sample while recording the local flux reading. This mode of operation is illustrated in Fig.~\ref{fig:setup}(c). This view typically shows vortices, with density that is determined by the magnetic field at which the ring was cooled. Figure~\ref{fig:scanned}(b) shows an image of such vortices, in units of $\Phi_0$. \par
(B) The local susceptibility was measured using a field coil (FC) placed around the pickup loop of the SQUID \cite{doi:10.1063/1.2932341}. The field generated by the field coil is eliminated by the superconductor, and this reduction is captured by the SQUID's pickup loop. Spatially mapping these signals yields a map of the diamagnetic susceptibility, presented in units of $\Phi_0/A$ (flux in the pick-up loop over FC current). Fig.~\ref{fig:setup}(d) demonstrates this mode of operation. Fig.~\ref{fig:scanned}(c) shows a representative susceptibility map, taken near the inner rim of the ring.\par
(C) The supercurrents flowing in the ring were imaged by mapping the flux generated by the current flow. An illustration is provided in Fig.~\ref{fig:setup}(e). The units are $\Phi_0/A$ (flux in the pick-up loop over EC current). A supercurrent map that was recorded at the same conditions as the susceptibility map is shown in Fig.~\ref{fig:scanned}(d).\par
Experiments (B) and (C) can be done either in dc or ac excitation currents. By performing both of them in ac mode at EC frequencies different than the FC frequencies we can separate the information in (A)-(C) and measure all three modes simultaneously.


In addition, we performed a global dc SQUID magnetization measurement with a pickup loop bigger than the entire SC ring. In this case the ring and EC move relative to the pickup loop, and the magnetic moment of the ring $M$ is determined as a function of EC current $I$ and temperature $T$   \cite{Keren_2022}.

\section{\label{sec:level1}Results}

\begin{figure}[h]
\includegraphics[width=0.49\textwidth]{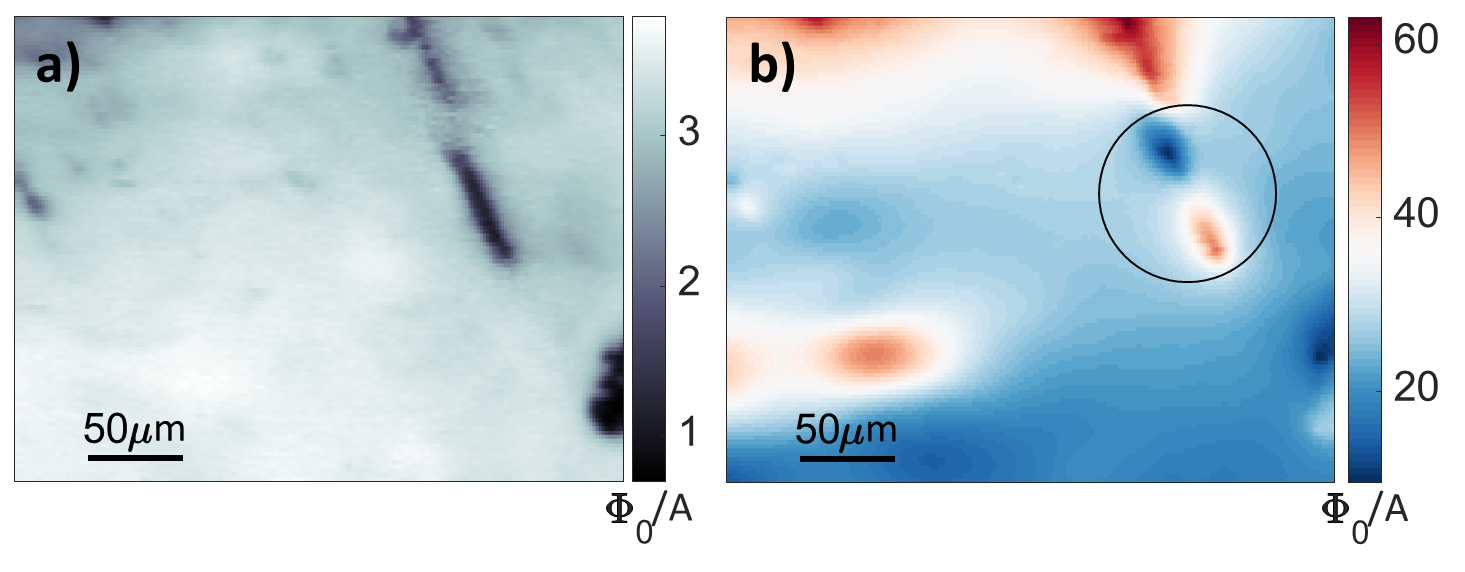}
\caption{\label{fig:flux_susc} \textbf{Local view of the destruction of superconductivity}. 
(a) Susceptibility map, showing the diamagnetic response of the sample to a current in the EC. Brighter regions represent stronger superconductivity. (b) Map of the flux generated by supercurrents flowing in the sample in response to the current in the EC. (a) and (b) were taken simultaneously in the presence of EC current of $I=0.05$~mA ac at $f_{ec}=223$~Hz, and field-coil current of $I=0.1$~mA at $f_{fc}=1237$~Hz. The supercurrents bend to bypass the region where superconductivity is destroyed, forming a dipole-like feature of reduced current flow, marked by a circle. This region appears as a dark streak in the susceptibility map.}
\end{figure}

 Figure~\ref{fig:flux_susc} shows a radial streak in the susceptibility map, along which the superconductivity is suppressed [Fig.~\ref{fig:flux_susc}(a)]. Figure~\ref{fig:flux_susc}(b) shows the magnetic fields generated by the supercurrents. The blue-red dipolelike feature indicates that the current is reduced in this region. This is consistent with the reduction of SC observed in the susceptibility map. The nucleation point for the streaks is determined by defects as locations with lower critical current, or by irregularities in the inner rim of the ring, which dictate in-homogeneity in the current flow resulting in a local increase in the current density.

\begin{figure}
\includegraphics[width=0.5\textwidth]{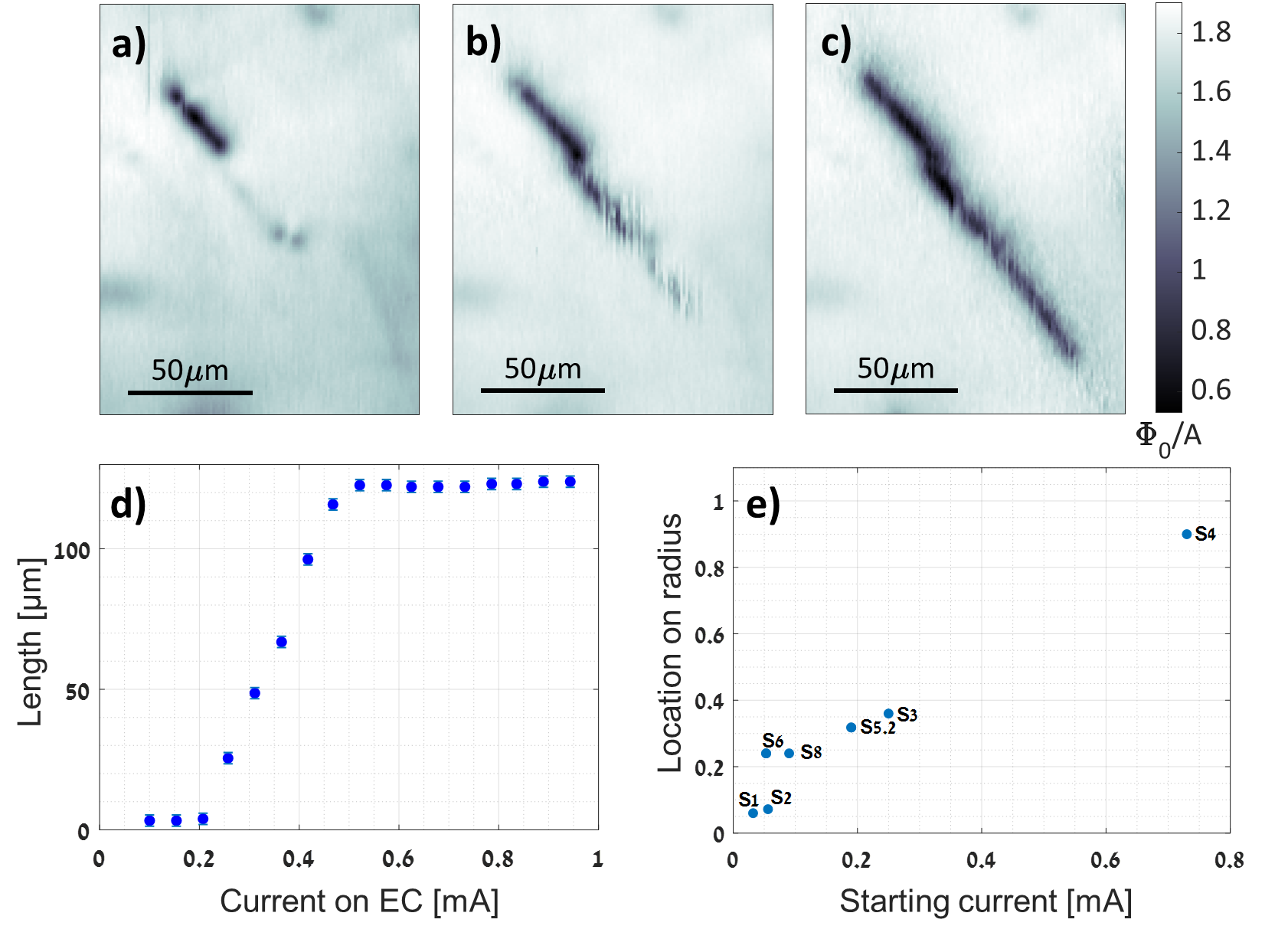}
\caption{\label{fig:streak_behavior} \textbf{Streak behavior}. (a)$-$(c) Susceptibility maps demonstrating the growth of a streak starting from a mid-ring nucleation point, at EC currents of $I=0.1$~mA, $I=0.41$~mA, and $I=0.52$~mA, respectively. (d) The length of a streak as a function of the EC current, demonstrating a threshold EC current, a linear growth, and a saturation after a certain EC current. (e) The location of nucleation points $S_i$ on the rings radius, as a function of the threshold current at which the streak starts to grow. Both the threshold current and saturation current (not shown) depend linearly on the distance from the ring center.}
\end{figure}

\begin{figure}
\includegraphics[width=0.4\textwidth]{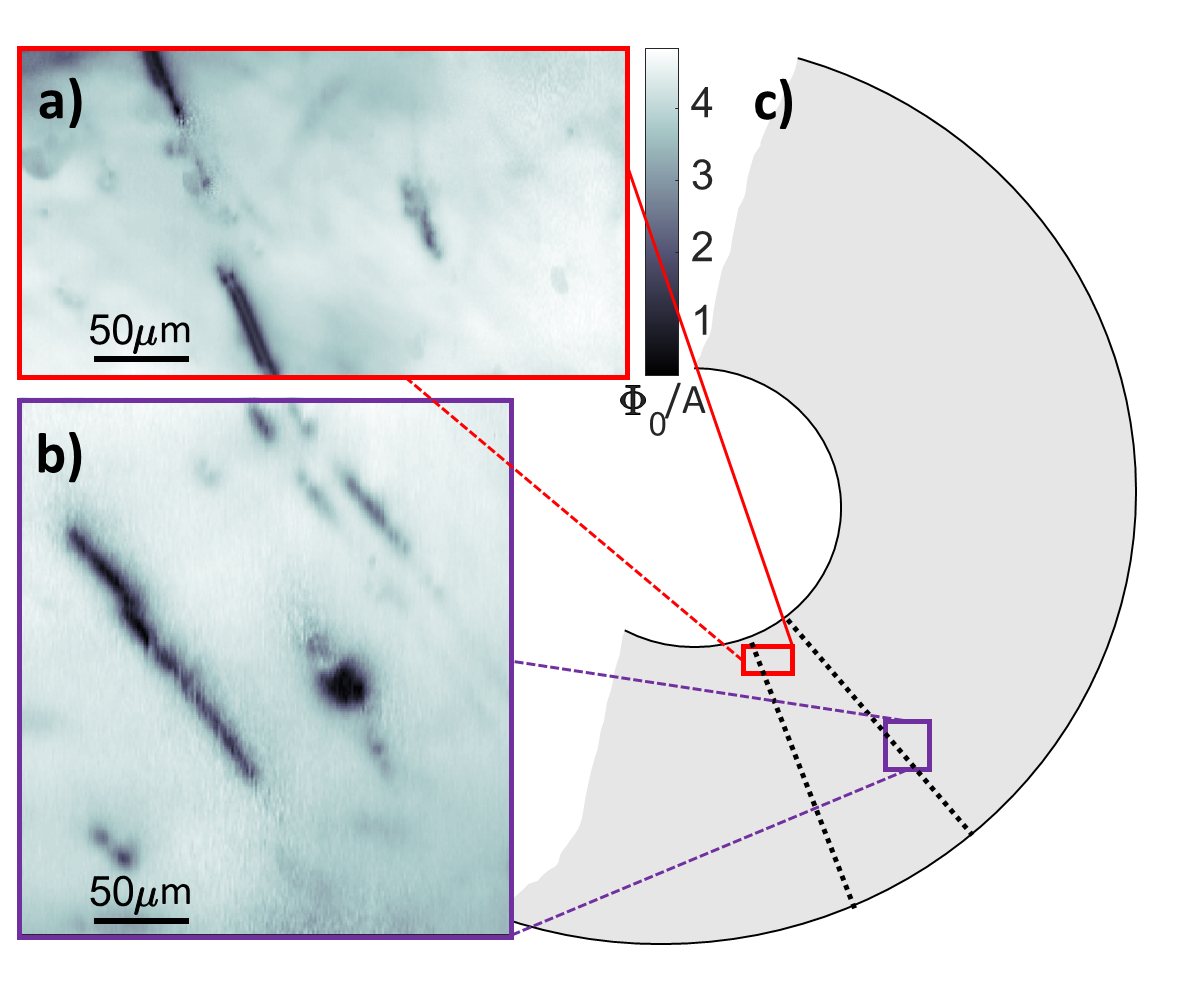}
\caption{\label{fig:radial_growth} \textbf{Radial growth of the streaks}. (a),(b) Susceptibility maps showing two streaks, at $I=0.04$~mA (a) and $I=0.52$~mA (b) flowing in the EC. The streak shown in (b) is located few 100's $\mu$m further out and at a different azimuth. (c) The location of streaks on the ring. Both streaks evolve along a radial lines (dotted lines) despite their location and azimuth.}
\end{figure}

The evolution of the normal streaks with $I$ can be followed by imaging susceptibility as a function of the current in the EC. A typical streak and its evolution is shown in Figs.~\ref{fig:streak_behavior}(a)$-$~\ref{fig:streak_behavior}(c). The streak is a line of reduced superconductivity, which progresses radially from the inner rim or from a defect on the ring itself. This streak would ideally grow linearly until it crosses the whole ring. In practice this streak grew linearly with applied current on the EC up to a certain value. After exceeding that streaks' critical current value the streak did not propagate further. The typical length of streaks in our sample was around 30 $\mu$m. This was the length of five out of eight streaks we followed. This is demonstrated in Fig.~\ref{fig:streak_behavior}(d) where we show the streak length as a function of $I$. The local decisions involved in the growth of each streak are controlled by the distribution of supercurrents. Somewhere along the ring another streak grows and navigates the supercurrents away from this region, hence the normal streak in Fig.~\ref{fig:streak_behavior}(c) no longer grows. 

Moreover, the location from which a streak starts to grow is also a linear function of $I$ as shown in Fig.~\ref{fig:streak_behavior}(e). Here we show the radius of a nucleation point in which a streak appears as a function of the EC current at which the streak first appears.

The radial nature of the streaks is independent of their distance from the inner rim on the ring. In Fig.~\ref{fig:radial_growth} we show two susceptibility images and the location where they where taken on the ring. Figure~\ref{fig:radial_growth}(a) is taken close to the inner rim at some azimuth and Fig.~\ref{fig:radial_growth}(b) is closer to the outer rim with a different azimuth. In both cases all the streaks, regardless of their nucleation point, are radial. 

\begin{figure}
\includegraphics[width=0.5\textwidth]{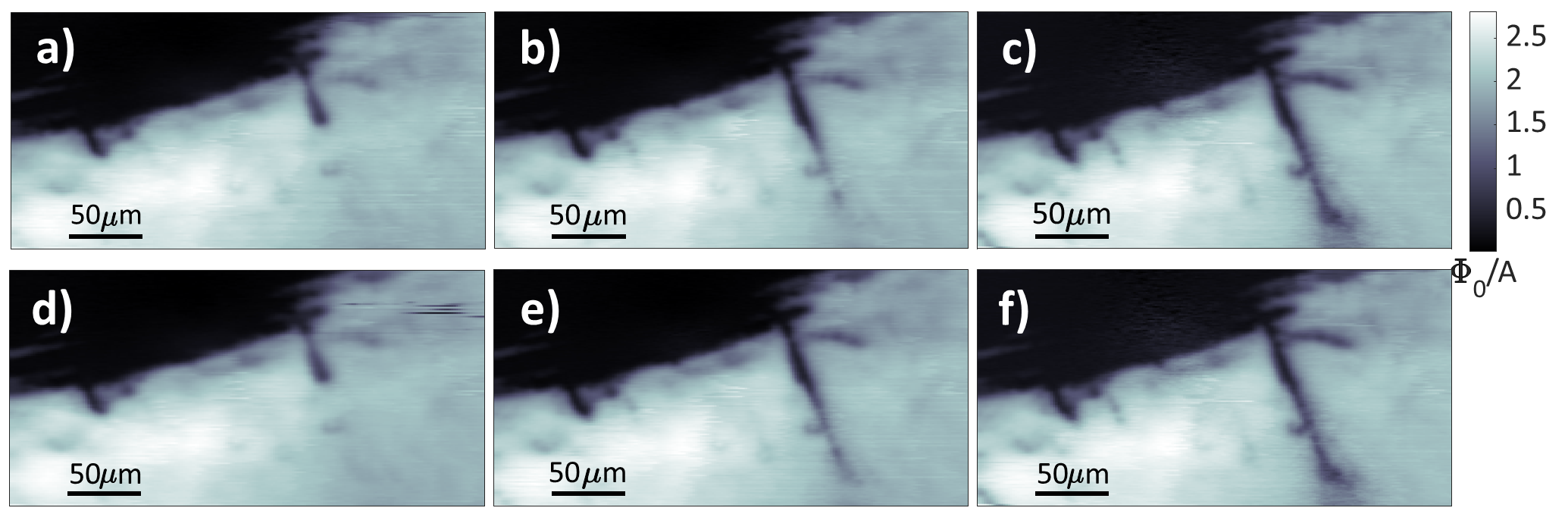}
\caption{\label{fig:streak_growth} \textbf{Reversibility of streaks and their propagation}.
(a)$-$(f) Susceptibility maps of a region near the inner rim, in the presence of EC current of (a) $I=0.003$~mA, (b) $I=0.054$~mA where a streak starts to grow, and (c) $I=0.093$~mA where the growth continued. Between (c) and (d) the current was turned off. (d) $I=0.003$~mA, demonstrating that the streak returned to its initial state. (e) $I=0.054$~mA, streak growth, and (f) $I=0.093$~ mA, the streak is back to the state it was in (c).}
\end{figure}

Finally, we demonstrate that the streaks are reversible. In Fig.~\ref{fig:streak_growth} we show the evolution of a streak that emerges from a nucleation site on the inner rim twice. First we followed the streak's growth [Figs.~\ref{fig:streak_growth}(a)$-$~\ref{fig:streak_growth}(c)], and then we turned off the current and observed a full recovery of the superconductivity at the location of the streak [Fig.~\ref{fig:streak_growth}(d)]. Increasing the current shows that the streak evolved in the same way [Figs.~\ref{fig:streak_growth}(e)$-$~\ref{fig:streak_growth}(f)]. In addition, in a different cooldown of the same sample we found that this streak nucleates at the same location (not shown). This demonstrates reversibility and reproducibility, both important for future applications. 
We note that the nucleation, growth, and direction of the streaks appear in both dc and ac currents in the EC.

\section{\label{sec:discussion}Discussion}

When a normal region is formed around a nucleation site due to the EC current, the spatial distribution of the supercurrents changes, in order to bypass this region [Fig.~\ref{fig:flux_susc}(b)]. When the EC current is further increased, the inhomogeneity of the current flow contributes to the growth of the streak normal regions, in addition to the natural disorder in the system. Since the overall direction of the current flow is around the ring, locations with maximum current, where the critical current is exceeded, progress radially along the ring (Fig.~\ref{fig:radial_growth}).

In our ring, several nucleation sites were found spread out at random locations on the ring's radius. All sites showed a characteristic behavior, namely, linear growth of the streak length [Fig.~\ref{fig:streak_behavior}(d)] with current, up to some maximum length. Nucleation sites further away from the center required higher currents on the EC to generate supercurrent flow that contributes to the growth of normal regions [Fig.~\ref{fig:streak_behavior}(e)]. In some cases the distance between two defects is only few tens of microns, making it possible in theory to connect them. 

The progress of a streak's front, emerging from the inner edge due to supercurrent flow, resembles dendrites as they both destroy SC in the sample. However, dendrites are usually generated by applying a magnetic field while streaks are generated through supercurrent flow around defects. Unlike dendrites  \cite{PhysRevB.52.75}, streaks are not pinned to the sample: They are reversible and they do not show the fractal shape. Control over dendrites is also limited; artificial triggering of instability that leads to dendrite growth does not result in the same branching patterns on repeated experimentation under the exact same conditions \cite{PhysRevLett.71.2646}. 

\begin{figure}
\includegraphics[width=0.45\textwidth]{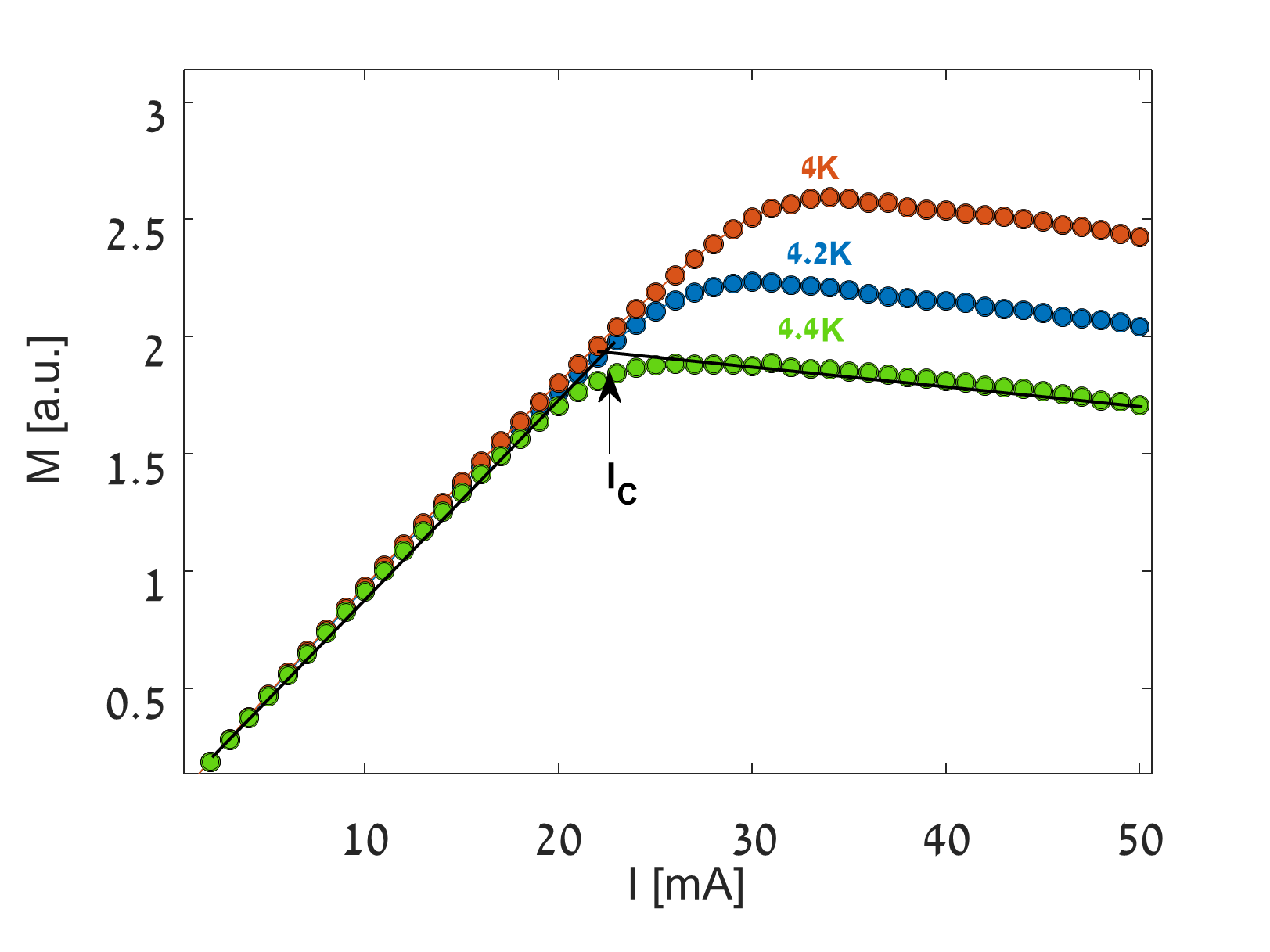}
\caption{\label{fig:I_and_T_dep} \textbf{Destruction of superconductivity of a whole ring}. Magnetic moments of a MoSi ring vs. current in EC. The figure shows a linear increase of the ring's magnetic moment up to a value of \(I_c\) where there is a knee. For \(I>I_c\) the moment decreases with increasing current as flux enters the sample. \(I_c\) increases with decreasing temperatures.}
\end{figure}

Using the inner EC we can generate a narrow non-superconducting path in the ring, with a radial orientation. For a narrow enough ring this might result in a ring with a slit that might behave as a Josephson junction \cite{magrini2019situ}. 
Looking at the global behavior of the system, the magnetization $M$ of the entire ring as a function of $I$ is depicted in Fig.~\ref{fig:I_and_T_dep}. Upon ramping $I$, $M$ increases linearly at first, but, for $I$ larger than some $I_c$, indicated on the figure, $M$ no longer increases with $I$, and even slightly decreases. This happens when flux finds a way to enter the ring along paths of normal state material, and according to Eq.~(\ref{JtoAandPhi}), reduce the current density in the ring. Naturally, $I_c$ decreases with increasing temperature as demonstrated in the figure. 

Back to the local view, high enough EC current destroyed superconductivity on radial slits in the ring. Reducing $I$ would return the superconductor to its fully SC state (no slit). Therefore, this potential Josephson junction might switch on and off, non-invasively, and reproducibly, resulting in a controllable device. The device can be switched from a ring with Josephson junctions to a fully SC ring, \textit{in situ}. By a careful design of nucleation points, it should be possible to fully determine the location and orientation of the slit(s). We therefore suggest that a SC ring with pre-designed fault points on the inner rim, pierced by a long EC, could act as a controllable SQUID where the normal region can be turned on and off at will.

Controllable Josephson junctions have been implemented in various ways, locally affecting the landscape of superconductivity. Among these are a reduction of electron density by electrostatic gating, destruction of superconductivity by proximity to a ferromagnet, and local heating with a focused laser beam \cite{magrini2019situ, lombardo2018situ, amado2013electrostatic, vavra2013current, begon2017high, goswami2016quantum, doi:10.1021/acs.nanolett.6b03820}. One example where a SQUID behavior was demonstrated in a reversible process is by electrostatic tuning of the local carrier density, in a two-dimensional superconductor that forms at the interface between the complex oxides lanthanum aluminate (LAO) and strontium titanate (STO) \cite{goswami2016quantum, doi:10.1021/acs.nanolett.6b03820, Kapon2019}. Each approach has advantages and disadvantages. 

\section{\label{sec:level1} Conclusions}

The main observation of this paper is a magnetic-field-free mixed state with radial and reproducible normal state streaks on a superconducting background. The streaks are produced by a long excitation coil piercing a ring-shaped SC, and are due to shielding currents which exceed locally the critical current. They emanate from defects, and cross the ring. This is an exciting feature looking for an application.

\section{Acknowledgments}

The authors thank Binghai Yan, Dganit Meidan, and Maxim Khodas for helpful discussions. All authors were supported by the Israeli Science Foundation, Grant No. ISF-1251/19. A. Khanukov and B. K. were also supported by the European Research Council Grant No. ERC-2019-COG-866236, the COST Action CA21144 and the Pazy Research Foundation. 

\begin{appendices}
 Current passing through an infinitely long coil creates a magnetic field only inside the coil. However, infinite coil is not a realistic option. In our case the EC is $7.5 \times 10^{6}$ longer than the SC height. While this is an impressive ratio, we want to check experimentally if it is good enough by measuring runaway fields.

\begin{figure}
\includegraphics[width=0.4\textwidth]{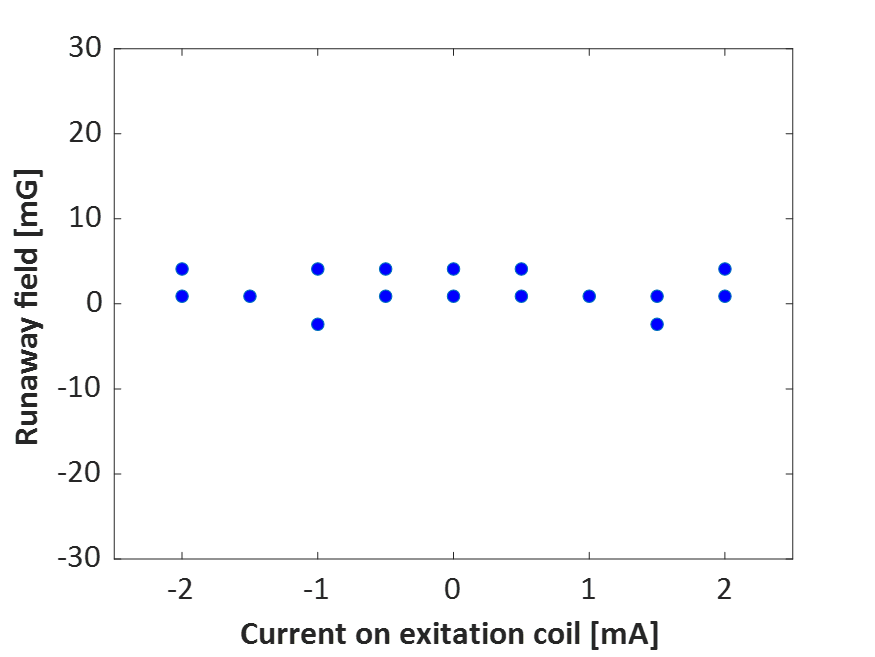}
\caption{\label{fig:runaway} \textbf{Runaway field from the long excitation coil}. Calculated runaway field from the number of vortices in the image, vs the current in the EC when ring is cooled down below $T_{c}$. The run-away field is measured by the number of trapped flux quanta in the ring [e.g., Fig.~\ref{fig:scanned}(a)]. Each point in the figure represents a calculated value of the field per cool down.}
\end{figure}

These fields can be determined by counting vortices trapped in the superconducting ring when it is cooled below $T_{c}$ in the presence of different currents on the EC. The conversion is made with the known relation
\begin{equation}
{\Phi} = \frac{ n \Phi_0}{A^2}
\end{equation} 
where $n$ is the number of vortices and $A^2$ is the scan area. 

Figure~\ref{fig:runaway} presents repeated measurements of the flux trapped by a SC ring, due to the EC current up to $0.6$~mG. No variation in the flux is detected when changing $I$. Therefore, the vortices are due to stray fields emanating from sources other than the coil. Moreover, we have measured the flux trapped in the sample in the presence of up to $20$~mA. Even under such high currents, which are ten times higher then the maximum applied in the experiment, no significant runaway fields were detected. Therefore, we can safely say that the EC is infinite for all practical purposes.

\end{appendices}

\nocite{*}

\bibliography{apssamp}

\providecommand{\noopsort}[1]{}\providecommand{\singleletter}[1]{#1}%
\begin{thebibliography}{14}%
\makeatletter
\providecommand \@ifxundefined [1]{%
 \@ifx{#1\undefined}
}%
\providecommand \@ifnum [1]{%
 \ifnum #1\expandafter \@firstoftwo
 \else \expandafter \@secondoftwo
 \fi
}%
\providecommand \@ifx [1]{%
 \ifx #1\expandafter \@firstoftwo
 \else \expandafter \@secondoftwo
 \fi
}%
\providecommand \natexlab [1]{#1}%
\providecommand \enquote  [1]{``#1''}%
\providecommand \bibnamefont  [1]{#1}%
\providecommand \bibfnamefont [1]{#1}%
\providecommand \citenamefont [1]{#1}%
\providecommand \href@noop [0]{\@secondoftwo}%
\providecommand \href [0]{\begingroup \@sanitize@url \@href}%
\providecommand \@href[1]{\@@startlink{#1}\@@href}%
\providecommand \@@href[1]{\endgroup#1\@@endlink}%
\providecommand \@sanitize@url [0]{\catcode `\\12\catcode `\$12\catcode
  `\&12\catcode `\#12\catcode `\^12\catcode `\_12\catcode `\%12\relax}%
\providecommand \@@startlink[1]{}%
\providecommand \@@endlink[0]{}%
\providecommand \url  [0]{\begingroup\@sanitize@url \@url }%
\providecommand \@url [1]{\endgroup\@href {#1}{\urlprefix }}%
\providecommand \urlprefix  [0]{URL }%
\providecommand \Eprint [0]{\href }%
\providecommand \doibase [0]{http://dx.doi.org/}%
\providecommand \selectlanguage [0]{\@gobble}%
\providecommand \bibinfo  [0]{\@secondoftwo}%
\providecommand \bibfield  [0]{\@secondoftwo}%
\providecommand \translation [1]{[#1]}%
\providecommand \BibitemOpen [0]{}%
\providecommand \bibitemStop [0]{}%
\providecommand \bibitemNoStop [0]{.\EOS\space}%
\providecommand \EOS [0]{\spacefactor3000\relax}%
\providecommand \BibitemShut  [1]{\csname bibitem#1\endcsname}%
\let\auto@bib@innerbib\@empty
\bibitem [{\citenamefont {Gavish}\ \emph {et~al.}(2021)\citenamefont {Gavish},
  \citenamefont {Kenneth},\ and\ \citenamefont {Keren}}]{GAVISH2021132767}%
  \BibitemOpen
  \bibfield  {author} {\bibinfo {author} {\bibfnamefont {N.}~\bibnamefont
  {Gavish}}, \bibinfo {author} {\bibfnamefont {O.}~\bibnamefont {Kenneth}}, \
  and\ \bibinfo {author} {\bibfnamefont {A.}~\bibnamefont {Keren}},\ }\href
  {\doibase https://doi.org/10.1016/j.physd.2020.132767} {\bibfield  {journal}
  {\bibinfo  {journal} {Physica D: Nonlinear Phenomena}\ }\textbf {\bibinfo
  {volume} {415}},\ \bibinfo {pages} {132767} (\bibinfo {year}
  {2021})}\BibitemShut {NoStop}%
\bibitem [{\citenamefont {Keren}\ \emph {et~al.}(2022)\citenamefont {Keren},
  \citenamefont {Blau}, \citenamefont {Gavish}, \citenamefont {Kenneth},
  \citenamefont {Ivry},\ and\ \citenamefont {Suleiman}}]{Keren_2022}%
  \BibitemOpen
  \bibfield  {author} {\bibinfo {author} {\bibfnamefont {A.}~\bibnamefont
  {Keren}}, \bibinfo {author} {\bibfnamefont {N.}~\bibnamefont {Blau}},
  \bibinfo {author} {\bibfnamefont {N.}~\bibnamefont {Gavish}}, \bibinfo
  {author} {\bibfnamefont {O.}~\bibnamefont {Kenneth}}, \bibinfo {author}
  {\bibfnamefont {Y.}~\bibnamefont {Ivry}}, \ and\ \bibinfo {author}
  {\bibfnamefont {M.}~\bibnamefont {Suleiman}},\ }\href {\doibase
  10.1088/1361-6668/ac7173} {\bibfield  {journal} {\bibinfo  {journal}
  {Superconductor Science and Technology}\ }\textbf {\bibinfo {volume} {35}},\
  \bibinfo {pages} {075013} (\bibinfo {year} {2022})}\BibitemShut {NoStop}%
\bibitem [{\citenamefont {Huber}\ \emph {et~al.}(2008)\citenamefont {Huber},
  \citenamefont {Koshnick}, \citenamefont {Bluhm}, \citenamefont {Archuleta},
  \citenamefont {Azua}, \citenamefont {Björnsson}, \citenamefont {Gardner},
  \citenamefont {Halloran}, \citenamefont {Lucero},\ and\ \citenamefont
  {Moler}}]{doi:10.1063/1.2932341}%
  \BibitemOpen
  \bibfield  {author} {\bibinfo {author} {\bibfnamefont {M.~E.}\ \bibnamefont
  {Huber}}, \bibinfo {author} {\bibfnamefont {N.~C.}\ \bibnamefont {Koshnick}},
  \bibinfo {author} {\bibfnamefont {H.}~\bibnamefont {Bluhm}}, \bibinfo
  {author} {\bibfnamefont {L.~J.}\ \bibnamefont {Archuleta}}, \bibinfo {author}
  {\bibfnamefont {T.}~\bibnamefont {Azua}}, \bibinfo {author} {\bibfnamefont
  {P.~G.}\ \bibnamefont {Björnsson}}, \bibinfo {author} {\bibfnamefont
  {B.~W.}\ \bibnamefont {Gardner}}, \bibinfo {author} {\bibfnamefont {S.~T.}\
  \bibnamefont {Halloran}}, \bibinfo {author} {\bibfnamefont {E.~A.}\
  \bibnamefont {Lucero}}, \ and\ \bibinfo {author} {\bibfnamefont {K.~A.}\
  \bibnamefont {Moler}},\ }\href {\doibase 10.1063/1.2932341} {\bibfield
  {journal} {\bibinfo  {journal} {Review of Scientific Instruments}\ }\textbf
  {\bibinfo {volume} {79}},\ \bibinfo {pages} {053704} (\bibinfo {year}
  {2008})},\ \Eprint {http://arxiv.org/abs/https://doi.org/10.1063/1.2932341}
  {https://doi.org/10.1063/1.2932341} \BibitemShut {NoStop}%
\bibitem [{\citenamefont {Persky}\ \emph {et~al.}(2022)\citenamefont {Persky},
  \citenamefont {Sochnikov},\ and\ \citenamefont
  {Kalisky}}]{doi:10.1146/annurev-conmatphys-031620-104226}%
  \BibitemOpen
  \bibfield  {author} {\bibinfo {author} {\bibfnamefont {E.}~\bibnamefont
  {Persky}}, \bibinfo {author} {\bibfnamefont {I.}~\bibnamefont {Sochnikov}}, \
  and\ \bibinfo {author} {\bibfnamefont {B.}~\bibnamefont {Kalisky}},\ }\href
  {\doibase 10.1146/annurev-conmatphys-031620-104226} {\bibfield  {journal}
  {\bibinfo  {journal} {Annual Review of Condensed Matter Physics}\ }\textbf
  {\bibinfo {volume} {13}},\ \bibinfo {pages} {null} (\bibinfo {year}
  {2022})},\ \Eprint
  {http://arxiv.org/abs/https://doi.org/10.1146/annurev-conmatphys-031620-104226}
  {https://doi.org/10.1146/annurev-conmatphys-031620-104226} \BibitemShut
  {NoStop}%
\bibitem [{\citenamefont {Dur\'an}\ \emph {et~al.}(1995)\citenamefont
  {Dur\'an}, \citenamefont {Gammel}, \citenamefont {Miller},\ and\
  \citenamefont {Bishop}}]{PhysRevB.52.75}%
  \BibitemOpen
  \bibfield  {author} {\bibinfo {author} {\bibfnamefont {C.~A.}\ \bibnamefont
  {Dur\'an}}, \bibinfo {author} {\bibfnamefont {P.~L.}\ \bibnamefont {Gammel}},
  \bibinfo {author} {\bibfnamefont {R.~E.}\ \bibnamefont {Miller}}, \ and\
  \bibinfo {author} {\bibfnamefont {D.~J.}\ \bibnamefont {Bishop}},\ }\href
  {\doibase 10.1103/PhysRevB.52.75} {\bibfield  {journal} {\bibinfo  {journal}
  {Phys. Rev. B}\ }\textbf {\bibinfo {volume} {52}},\ \bibinfo {pages} {75}
  (\bibinfo {year} {1995})}\BibitemShut {NoStop}%
\bibitem [{\citenamefont {Leiderer}\ \emph {et~al.}(1993)\citenamefont
  {Leiderer}, \citenamefont {Boneberg}, \citenamefont {Br\"ull}, \citenamefont
  {Bujok},\ and\ \citenamefont {Herminghaus}}]{PhysRevLett.71.2646}%
  \BibitemOpen
  \bibfield  {author} {\bibinfo {author} {\bibfnamefont {P.}~\bibnamefont
  {Leiderer}}, \bibinfo {author} {\bibfnamefont {J.}~\bibnamefont {Boneberg}},
  \bibinfo {author} {\bibfnamefont {P.}~\bibnamefont {Br\"ull}}, \bibinfo
  {author} {\bibfnamefont {V.}~\bibnamefont {Bujok}}, \ and\ \bibinfo {author}
  {\bibfnamefont {S.}~\bibnamefont {Herminghaus}},\ }\href {\doibase
  10.1103/PhysRevLett.71.2646} {\bibfield  {journal} {\bibinfo  {journal}
  {Phys. Rev. Lett.}\ }\textbf {\bibinfo {volume} {71}},\ \bibinfo {pages}
  {2646} (\bibinfo {year} {1993})}\BibitemShut {NoStop}%
\bibitem [{\citenamefont {Magrini}\ \emph {et~al.}(2019)\citenamefont
  {Magrini}, \citenamefont {Mironov}, \citenamefont {Rochet}, \citenamefont
  {Tamarat}, \citenamefont {Buzdin},\ and\ \citenamefont
  {Lounis}}]{magrini2019situ}%
  \BibitemOpen
  \bibfield  {author} {\bibinfo {author} {\bibfnamefont {W.}~\bibnamefont
  {Magrini}}, \bibinfo {author} {\bibfnamefont {S.}~\bibnamefont {Mironov}},
  \bibinfo {author} {\bibfnamefont {A.}~\bibnamefont {Rochet}}, \bibinfo
  {author} {\bibfnamefont {P.}~\bibnamefont {Tamarat}}, \bibinfo {author}
  {\bibfnamefont {A.~I.}\ \bibnamefont {Buzdin}}, \ and\ \bibinfo {author}
  {\bibfnamefont {B.}~\bibnamefont {Lounis}},\ }\href@noop {} {\bibfield
  {journal} {\bibinfo  {journal} {Applied Physics Letters}\ }\textbf {\bibinfo
  {volume} {114}},\ \bibinfo {pages} {142601} (\bibinfo {year}
  {2019})}\BibitemShut {NoStop}%
\bibitem [{\citenamefont {Lombardo}\ \emph {et~al.}(2018)\citenamefont
  {Lombardo}, \citenamefont {Jeli{\'c}}, \citenamefont {Baumans}, \citenamefont
  {Scheerder}, \citenamefont {Nacenta}, \citenamefont {Moshchalkov},
  \citenamefont {Van~de Vondel}, \citenamefont {Kramer}, \citenamefont
  {Milo{\v{s}}evi{\'c}},\ and\ \citenamefont {Silhanek}}]{lombardo2018situ}%
  \BibitemOpen
  \bibfield  {author} {\bibinfo {author} {\bibfnamefont {J.}~\bibnamefont
  {Lombardo}}, \bibinfo {author} {\bibfnamefont {{\v{Z}}.~L.}\ \bibnamefont
  {Jeli{\'c}}}, \bibinfo {author} {\bibfnamefont {X.~D.}\ \bibnamefont
  {Baumans}}, \bibinfo {author} {\bibfnamefont {J.~E.}\ \bibnamefont
  {Scheerder}}, \bibinfo {author} {\bibfnamefont {J.~P.}\ \bibnamefont
  {Nacenta}}, \bibinfo {author} {\bibfnamefont {V.~V.}\ \bibnamefont
  {Moshchalkov}}, \bibinfo {author} {\bibfnamefont {J.}~\bibnamefont {Van~de
  Vondel}}, \bibinfo {author} {\bibfnamefont {R.~B.}\ \bibnamefont {Kramer}},
  \bibinfo {author} {\bibfnamefont {M.~V.}\ \bibnamefont
  {Milo{\v{s}}evi{\'c}}}, \ and\ \bibinfo {author} {\bibfnamefont {A.~V.}\
  \bibnamefont {Silhanek}},\ }\href@noop {} {\bibfield  {journal} {\bibinfo
  {journal} {Nanoscale}\ }\textbf {\bibinfo {volume} {10}},\ \bibinfo {pages}
  {1987} (\bibinfo {year} {2018})}\BibitemShut {NoStop}%
\bibitem [{\citenamefont {Amado}\ \emph {et~al.}(2013)\citenamefont {Amado},
  \citenamefont {Fornieri}, \citenamefont {Carillo}, \citenamefont {Biasiol},
  \citenamefont {Sorba}, \citenamefont {Pellegrini},\ and\ \citenamefont
  {Giazotto}}]{amado2013electrostatic}%
  \BibitemOpen
  \bibfield  {author} {\bibinfo {author} {\bibfnamefont {M.}~\bibnamefont
  {Amado}}, \bibinfo {author} {\bibfnamefont {A.}~\bibnamefont {Fornieri}},
  \bibinfo {author} {\bibfnamefont {F.}~\bibnamefont {Carillo}}, \bibinfo
  {author} {\bibfnamefont {G.}~\bibnamefont {Biasiol}}, \bibinfo {author}
  {\bibfnamefont {L.}~\bibnamefont {Sorba}}, \bibinfo {author} {\bibfnamefont
  {V.}~\bibnamefont {Pellegrini}}, \ and\ \bibinfo {author} {\bibfnamefont
  {F.}~\bibnamefont {Giazotto}},\ }\href@noop {} {\bibfield  {journal}
  {\bibinfo  {journal} {Physical Review B}\ }\textbf {\bibinfo {volume} {87}},\
  \bibinfo {pages} {134506} (\bibinfo {year} {2013})}\BibitemShut {NoStop}%
\bibitem [{\citenamefont {Vavra}\ \emph {et~al.}(2013)\citenamefont {Vavra},
  \citenamefont {Pfaff}, \citenamefont {Monaco}, \citenamefont {Aprili},\ and\
  \citenamefont {Strunk}}]{vavra2013current}%
  \BibitemOpen
  \bibfield  {author} {\bibinfo {author} {\bibfnamefont {O.}~\bibnamefont
  {Vavra}}, \bibinfo {author} {\bibfnamefont {W.}~\bibnamefont {Pfaff}},
  \bibinfo {author} {\bibfnamefont {R.}~\bibnamefont {Monaco}}, \bibinfo
  {author} {\bibfnamefont {M.}~\bibnamefont {Aprili}}, \ and\ \bibinfo {author}
  {\bibfnamefont {C.}~\bibnamefont {Strunk}},\ }\href@noop {} {\bibfield
  {journal} {\bibinfo  {journal} {Applied Physics Letters}\ }\textbf {\bibinfo
  {volume} {102}},\ \bibinfo {pages} {072602} (\bibinfo {year}
  {2013})}\BibitemShut {NoStop}%
\bibitem [{\citenamefont {B{\'e}gon-Lours}\ \emph {et~al.}(2017)\citenamefont
  {B{\'e}gon-Lours}, \citenamefont {Rouco}, \citenamefont {Sander},
  \citenamefont {Trastoy}, \citenamefont {Bernard}, \citenamefont {Jacquet},
  \citenamefont {Bouzehouane}, \citenamefont {Fusil}, \citenamefont {Garcia},
  \citenamefont {Barth{\'e}l{\'e}my} \emph {et~al.}}]{begon2017high}%
  \BibitemOpen
  \bibfield  {author} {\bibinfo {author} {\bibfnamefont {L.}~\bibnamefont
  {B{\'e}gon-Lours}}, \bibinfo {author} {\bibfnamefont {V.}~\bibnamefont
  {Rouco}}, \bibinfo {author} {\bibfnamefont {A.}~\bibnamefont {Sander}},
  \bibinfo {author} {\bibfnamefont {J.}~\bibnamefont {Trastoy}}, \bibinfo
  {author} {\bibfnamefont {R.}~\bibnamefont {Bernard}}, \bibinfo {author}
  {\bibfnamefont {E.}~\bibnamefont {Jacquet}}, \bibinfo {author} {\bibfnamefont
  {K.}~\bibnamefont {Bouzehouane}}, \bibinfo {author} {\bibfnamefont
  {S.}~\bibnamefont {Fusil}}, \bibinfo {author} {\bibfnamefont
  {V.}~\bibnamefont {Garcia}}, \bibinfo {author} {\bibfnamefont
  {A.}~\bibnamefont {Barth{\'e}l{\'e}my}},  \emph {et~al.},\ }\href@noop {}
  {\bibfield  {journal} {\bibinfo  {journal} {Physical Review Applied}\
  }\textbf {\bibinfo {volume} {7}},\ \bibinfo {pages} {064015} (\bibinfo {year}
  {2017})}\BibitemShut {NoStop}%
\bibitem [{\citenamefont {Goswami}\ \emph {et~al.}(2016)\citenamefont
  {Goswami}, \citenamefont {Mulazimoglu}, \citenamefont {Monteiro},
  \citenamefont {W{\"o}lbing}, \citenamefont {Koelle}, \citenamefont {Kleiner},
  \citenamefont {Blanter}, \citenamefont {Vandersypen},\ and\ \citenamefont
  {Caviglia}}]{goswami2016quantum}%
  \BibitemOpen
  \bibfield  {author} {\bibinfo {author} {\bibfnamefont {S.}~\bibnamefont
  {Goswami}}, \bibinfo {author} {\bibfnamefont {E.}~\bibnamefont
  {Mulazimoglu}}, \bibinfo {author} {\bibfnamefont {A.~M.}\ \bibnamefont
  {Monteiro}}, \bibinfo {author} {\bibfnamefont {R.}~\bibnamefont
  {W{\"o}lbing}}, \bibinfo {author} {\bibfnamefont {D.}~\bibnamefont {Koelle}},
  \bibinfo {author} {\bibfnamefont {R.}~\bibnamefont {Kleiner}}, \bibinfo
  {author} {\bibfnamefont {Y.~M.}\ \bibnamefont {Blanter}}, \bibinfo {author}
  {\bibfnamefont {L.~M.}\ \bibnamefont {Vandersypen}}, \ and\ \bibinfo {author}
  {\bibfnamefont {A.~D.}\ \bibnamefont {Caviglia}},\ }\href@noop {} {\bibfield
  {journal} {\bibinfo  {journal} {Nature nanotechnology}\ }\textbf {\bibinfo
  {volume} {11}},\ \bibinfo {pages} {861} (\bibinfo {year} {2016})}\BibitemShut
  {NoStop}%
\bibitem [{\citenamefont {Monteiro}\ \emph {et~al.}(2017)\citenamefont
  {Monteiro}, \citenamefont {Groenendijk}, \citenamefont {Manca}, \citenamefont
  {Mulazimoglu}, \citenamefont {Goswami}, \citenamefont {Blanter},
  \citenamefont {Vandersypen},\ and\ \citenamefont
  {Caviglia}}]{doi:10.1021/acs.nanolett.6b03820}%
  \BibitemOpen
  \bibfield  {author} {\bibinfo {author} {\bibfnamefont {A.~M. R. V.~L.}\
  \bibnamefont {Monteiro}}, \bibinfo {author} {\bibfnamefont {D.~J.}\
  \bibnamefont {Groenendijk}}, \bibinfo {author} {\bibfnamefont
  {N.}~\bibnamefont {Manca}}, \bibinfo {author} {\bibfnamefont
  {E.}~\bibnamefont {Mulazimoglu}}, \bibinfo {author} {\bibfnamefont
  {S.}~\bibnamefont {Goswami}}, \bibinfo {author} {\bibfnamefont
  {Y.}~\bibnamefont {Blanter}}, \bibinfo {author} {\bibfnamefont {L.~M.~K.}\
  \bibnamefont {Vandersypen}}, \ and\ \bibinfo {author} {\bibfnamefont {A.~D.}\
  \bibnamefont {Caviglia}},\ }\href {\doibase 10.1021/acs.nanolett.6b03820}
  {\bibfield  {journal} {\bibinfo  {journal} {Nano Letters}\ }\textbf {\bibinfo
  {volume} {17}},\ \bibinfo {pages} {715} (\bibinfo {year} {2017})},\ \bibinfo
  {note} {pMID: 28071920},\ \Eprint
  {http://arxiv.org/abs/https://doi.org/10.1021/acs.nanolett.6b03820}
  {https://doi.org/10.1021/acs.nanolett.6b03820} \BibitemShut {NoStop}%
\bibitem [{\citenamefont {Kapon}\ \emph {et~al.}(2019)\citenamefont {Kapon},
  \citenamefont {Salman}, \citenamefont {Mangel}, \citenamefont {Prokscha},
  \citenamefont {Gavish},\ and\ \citenamefont {Keren}}]{Kapon2019}%
  \BibitemOpen
  \bibfield  {author} {\bibinfo {author} {\bibfnamefont {I.}~\bibnamefont
  {Kapon}}, \bibinfo {author} {\bibfnamefont {Z.}~\bibnamefont {Salman}},
  \bibinfo {author} {\bibfnamefont {I.}~\bibnamefont {Mangel}}, \bibinfo
  {author} {\bibfnamefont {T.}~\bibnamefont {Prokscha}}, \bibinfo {author}
  {\bibfnamefont {N.}~\bibnamefont {Gavish}}, \ and\ \bibinfo {author}
  {\bibfnamefont {A.}~\bibnamefont {Keren}},\ }\href {\doibase
  10.1038/s41467-019-10480-x} {\bibfield  {journal} {\bibinfo  {journal}
  {Nature Communications}\ }\textbf {\bibinfo {volume} {10}},\ \bibinfo {pages}
  {2463} (\bibinfo {year} {2019})}\BibitemShut {NoStop}%
\end{thebibliography}%

\end{document}